\documentclass[12pt]{iopart}
\expandafter\let\csname equation*\endcsname\relax
\expandafter\let\csname endequation*\endcsname\relax
\usepackage{amssymb,amsmath,amsfonts,bbold,dsfont}
\usepackage{cite}
\begin{document}
\title[Dynamic localization in an effective tight binding Hamiltonian model ...] {Dynamic localization in an effective tight binding Hamiltonian model with a rapidly oscillating homogeneous electric field on a lattice}

\author{L. A. Mart{\'i}nez - Quintana}
\address{Departamento de F{\'i}sica, Universidad Sim\'on Bol{\'i}var, Apartado 89000, Caracas 1080A, Venezuela.}

\author{L. A. Gonz\'alez - D{\'i}az}
\address{Laboratorio de F{\'i}sica Estad{\'i}stica de Medios Desordenados. Centro de F\'{\i}sica, Instituto Venezolano de Investigaciones Cient\'{\i}ficas, Caracas 1020 - A, Venezuela.}
\ead{lagdivic@gmail.com}
\vspace{10pt}

\begin{abstract}
By the Magnus-Floquet approach we calculate the effective Hamiltonian for a charged particle on the lattice subject to a homogeneous high frequency oscillating electric field. The obtained result indicate the absence of dynamic localization of the particle for any value of the lattice constant and  electric field applied, which completes the limit results obtained by Dunlap and Kenkre.
\end{abstract}

%
\vspace{2pc}

%
\section{Introduction}
Dynamic localization is a versatile quantum phenomena of electrons widely studied by its potential technological applications  \cite{Dunlap,Zhao,Holthaus96,Papp,Papp2,Wang2015}. On the other hand,
 recent experiments show the analogy between photonics wave packets and the Bloch electrons on the lattice subject to electromagnetic fields\cite{Longhi2007,Dreisow2011}. Based on this evidence, it has been possible to construct optical lattices, where the dynamic localization of photonic wave packets has been observed experimentally \cite{Longhi2006}. Optical fibers arrays are the most direct application of these quantum systems \cite{Garanovich2012,Georgescu2014}. 

It is known that in the free space, i.e., in the absence of a lattice, high frequency electromagnetic fields produce effects of confinements that trap particles in demarcated regions of space \cite{Rahav,Bandyopadhyay2008}. In the literature, the theoretical approach to study the dynamics of these systems is through the calculation of effective Hamiltonians. This technique has been used with great success in areas such as  high resolution NMR spectroscopy \cite{Maricq}, the topological states of Floquet in graphene and semi-metals under the action of external electromagnetic fields \cite{Cayssol2013,Lindher2011,Calvo2011,Wang2013,Delplace2013,Grushin2014,GomezLeon2014,Titum,Quelle}, in superfluid systems \cite{Jiang2011,Liu2012,Tong2013,Thakurathi2013}, among others. It has also been applied the approach of effective Hamiltonians in systems discretized spatially by the presence of the lattice \cite{Itin2014,Martinez,Eckardt}. 

In the calculation of effective Hamiltonians at least two different models are used, namely: the method of multi-temporal scales \cite{Rahav} and the Magnus-Floquet method \cite{Maricq}. The purpose of this work is to calculate the tight binding effective Hamiltonian, $G$, of a charged particle in a one dimensional lattice subject to a homogeneous electric field dependent on time, periodic and high frequency. We will be used the approach of Magnus-Floquet for the calculation of $G$ and to study if the dynamic localization effects predicted by Dunlap and Kenkre \cite{Dunlap} remain at high frequencies. We find, as main result, that at high frequency the particle does not present dynamic localization. Starting from the results of Dunlap y Kenkre, we can not obtain the behavior of the mean square displacement at high frequency. It is necessary to apply the formalism of Magnus-Floquet in order to obtain the behavior at high frequency. Experiments in optical lattices would allow to verify this theoretical prediction obtained.
    
\section{Effective Hamiltonian for periodically driven system.}
Floquet's theorem \cite{Floquet,Gesztesy,Levante,Nauts,Casas,Blanes,Maricq,Rahav,Goldman} has been applied to time-dependent quantum systems in two ways: (a) The time-dependent propagation technique exploits that the propagator $U(t)$ for a periodic Hamiltonian $H(t)= H(t + nT)$ ($n\in\mathbb{Z}$) is of the form $U(t) = P(t)\,e^{-i\,G\,t}$, where $G$ is a time-independent Hamiltonian and $P(t)$ is cyclic: $P(t) = P(t + nT)$. And (b) in the time-independent Floquet Hamiltonian approach \cite{Sambe,Kavanaugh,Eckardt,Eckardt2}, the time-dependent Hilbert-space Hamiltonian is transformed into an infinite-dimensional Hilbert space, the so-called Floquet space, where it is represented by a time-independent Floquet Hamiltonian. The time-dependent problem reduces to a time-independent one, but with infinite dimension. In this paper, we will follow the first way.

We consider the coherent hopping dynamics of a quantum particle on a one-dimensional tight-binding lattice rapidly driven by an external sinusoidal field with homogeneous hopping rates, which is described by the Hamiltonian (with $\hslash=1$ and $e=1$)
\begin{equation}\label{hamiltonianot}
H(t)=H_{0}+H_{1}(t)
\end{equation}
with
\begin{equation}\label{hamiltoniano0}
H_{0}=-2A\cos(a p)
\end{equation}
and
\begin{equation}\label{hamiltoniano1}
H_{1}(t)=\epsilon x \cos(\omega t)
\end{equation}
$A$ is the tight-binding matrix element for the hopping of the particle, $a$ is the lattice constant and $\epsilon$ is a perturbing force. Here, we will consider that $\omega\gg\omega_{B}$, with $\omega_{B}=\epsilon a$ the Bloch's frequency.
%
%

Starting from the approach of Maricq \cite{Maricq}, we have that
\begin{equation}\label{g}
G=\sum_{k=0}G^{(k)}
\end{equation}
with
\begin{equation}\label{gs}
G^{(k)}=\frac{1}{T}\int_{0}^{T}\{H(t^{\prime})P^{(k)}(t^{\prime})-\sum_{j=1}^{k-1}P^{(j)}(t)\,G^{(k-j)}\}\,dt^{\prime},
\end{equation}
$G^{(0)}=0$, where
\begin{equation}\label{p}
P(t)=\sum_{k=0}P^{(k)}(t)
\end{equation}
with
\begin{equation}\label{ps}
P^{(k)}(t)=-i\int_{0}^{t}\{H(t^{\prime})P^{(k-1)}(t^{\prime})-\sum_{j=1}^{k-1}P^{(j)}(t)\,G^{(k-j)}-G^{(k)}\}\,dt^{\prime}.
\end{equation}
$P^{(0)}(t)=\mathds{1}$. For the Hamiltonian \eqref{hamiltonianot}, we have that
\begin{align}\label{g1}
G^{(1)}&=\frac{1}{T}\int_{0}^{T}H(t^{\prime})\,dt^{\prime}\\\nonumber
&=H_{0},
\end{align}

\begin{align}\label{p1}
P^{(1)}(t)&=-i \int_{0}^{t}\left(H(t^{\prime})-G^{(1)}\right)\,dt^{\prime}\\\nonumber
&=\epsilon \frac{\sin(\omega t)}{\omega}\frac{\partial}{\partial p},
\end{align}

\begin{align}\label{g2}
G^{(2)}&=\frac{1}{T}\int_{0}^{T}
\left(\left[H_{0},P^{(1)}(t^{\prime})\right]+H_{1}(t^{\prime})\right)\,dt^{\prime}\\\nonumber
&=0,
\end{align}

\begin{align}\label{p2}
P^{(2)}(t)&=-i\int_{0}^{t}
\left(\left[H_{0},P^{(1)}(t^{\prime})\right]+H_{1}(t^{\prime})P^{(1)}(t^{\prime})\right)\,dt^{\prime}\\\nonumber
&=\frac{\epsilon}{\omega^{2}}\sin^2\left(\frac{\omega t}{2}\right)\left(i\,4Aa\sin(a p)+2\epsilon\cos^2\left(\frac{\omega t}{2}\right)\frac{\partial^2}{\partial p^2}\right),
\end{align}

\begin{align}\label{g3}
G^{(3)}&=\frac{1}{T}\int_{0}^{T}
\left(\left[H_{0},P^{(2)}(t^{\prime})\right]+H_{1}(t^{\prime})P^{(2)}(t^{\prime})\right)\,dt^{\prime}\\\nonumber
&=\frac{Aa^2\epsilon^2}{2\omega^2}\cos(a p)
\end{align}



From \eqref{g}, the time-independent Hamiltonian is
\begin{equation}\label{hamilefectivo}
G\approx G^{(1)}+G^{(2)}+G^{(3)}=-2A\left(1-\frac{a^{2}\epsilon^{2}}{4\,\omega^{2}}\right)\cos\left(a p\right)=\mathcal{J}H_{0}
\end{equation}
with $\mathcal{J}\equiv 1-\frac{a^{2}\epsilon^{2}}{4\,\omega^{2}}$, which coincides with the first two terms of the Bessel function $J_{0}$. The result obtained in \eqref{hamilefectivo} is consistent up to $O(\omega^{-2})$ with the exact result for the band narrowing in the presence of a homogeneous oscillating electric field \cite{Dunlap}. The band narrowing effect has been observed experimentally with cold atoms in optical lattices \cite{Lignier}. From \eqref{p}, we have that
\begin{align}\label{pe}
P(t)\approx P^{(0)}(t)&+P^{(1)}(t)+P^{(2)}(t)=\mathds{1}+\epsilon \frac{\sin(\omega t)}{\omega}\frac{\partial}{\partial p}+\\\nonumber
&\frac{\epsilon}{\omega^{2}}\sin^2\left(\frac{\omega t}{2}\right)\left(i\,4Aa\sin(a p)+2\epsilon\cos^2\left(\frac{\omega t}{2}\right)\frac{\partial^2}{\partial p^2}\right)
\end{align}
The wave function is given by
\begin{equation}\label{funcion-onda}
\Psi(t,p)=P(t)\,e^{-i\,t\,G}\,\psi(p)
\end{equation}
with $\psi(p)$ a generalized eigenfunction of $G$.

Dynamic localization can be seen through the mean square displacement, which for $t\gg\frac{2\pi}{\omega}$ is given by
\begin{align}\label{vcm}
\langle\Psi|X^2|\Psi\rangle=&-\int_{-\infty}^{\infty}\Psi^{\ast}(t,p)\,\frac{\partial^2}{\partial p^2}\Psi(t,p)\,dp
\sim  -2A^2\biggl(a^2\left(1-\frac{1}{2}\left(\frac{\omega_{B}}{\omega}\right)^2\right) I_{1} +\\\nonumber &
+a\sin\left(\omega t \right)\left(\frac{\omega_{B}}{\omega}\right)\tilde{I_{1}}+\sin^2\left(\omega t\right) \left(\frac{\omega_{B}}{\omega}\right)^2\left(\tilde{I_{2}}+2a^2I_{2} \right)\biggr)t^2
\end{align}
with 
\begin{align*}
&I_{1}\equiv \int_{-\infty}^{\infty}\sin^2(a p)\psi^{\ast}(p)\,\psi(p)\,dp \\
&I_{2}\equiv  \int_{-\infty}^{\infty}\cos(2a p)\psi^{\ast}(p)\,\psi(p)\,dp\\
&\tilde{I_{1}}\equiv \sin^2(a p) \psi^{\ast}(p)\,\psi(p)\,\Big|_{-\infty}^{\infty}\\   
&\tilde{I_{2}}\equiv \sin^2(a p) \frac{d}{dp} (\psi^{\ast}(p)\,\psi(p))\,\Big|_{-\infty}^{\infty}
\end{align*}
The mean square displacement is not bounded in time, which physically means that the particle does not exhibit dynamic localization.

The dynamic localization condition of a charged particle in the presence of an oscillating and homogeneous electric field is deduced from the exact formulae of the mean square displacement obtained by Dunlap and Kenkre \cite{Dunlap}. From this formulae, it follows that if  $\frac{a\epsilon}{\omega}$ is greater or less than a root of the Bessel function $J_{0}$ the dynamic localization phenomenon disappears. In the context of high frequency, \eqref{vcm} shows the dependence of time of the mean square displacement, specifically when $t\gg\frac{2\pi}{\omega}$ (larges times). The high frequency limit can not be taken from the exact result offered by Dunlap and Kenkre, because the functions found in the Dunlap and Kenkre paper \cite{Dunlap} $A_{u}$ and $A_{v}$ are not defined in that limit. Therefore, the Magnus-Floquet approach is necessary to obtain the dynamic behavior of the particle on the lattice.

\section{Conclusions}
Based on the perturbative formalism introduced in \cite{Maricq}, one obtains an average Hamiltonian equivalent to that obtained from the Magnus expansion. This formalism is very practical to calculate the effective Hamiltonian $G$ on the lattice in the tight binding approximation. In this sense, we obtained $G$ up to $O(\omega^{-2})$ of charged particle on a lattice subject to a homogeneous and periodically oscillating electric field at high frequency. The effective Hamiltonian exhibits the well-known renormalization of the hopping rate by a Bessel function of the first kind, $J_{0}$. This renormalization is known as band narrowing effect \cite{Dunlap}. It should be mentioned that this effect has also been obtained by semi-classical model \cite{Martinez}. Following the time-independent Floquet Hamiltonian approach, we obtained the wave-function $\Psi(t,p)$ associated to $H(t)$ up to $O(\omega^{-2})$. Finally, we calculated the mean square displacement for larges time, noting that it is not bounded in time, indicating that the particle does exhibit dynamic localization, independently of the values of $a$ and $\epsilon$. By virtue of this result, it could be stated that the effective electric field is inhomogeneous since the Bloch oscillations are not observed. The result obtained completes the limit results shown by Dunlap and Kenkre.

\section{Acknowledgments}
This work was supported by IVIC under Project No. 1089.
\\

\end{document}